# Coherent microscopy by laser optical feedback imaging (LOFI) technique

O. Hugon, F. Joud[1], E. Lacot, O. Jacquin, H. Guillet de Chatellus.

UJF - Grenoble 1 / CNRS, LIPhy UMR 5588, Grenoble, F-38041, France.

*Corresponding author*:   Olivier HUGON

Laboratoire interdisciplinaire de Physique

B.P. 87

38402 Saint Martin d'Hères Cedex FRANCE

Tel: (33)+4 76 51 47 49

Fax: (33)+4 76 63 54 95

E-mail: olivier.hugon@ujf-grenoble.fr

## *Abstract*

The application of the non conventional imaging technique LOFI (Laser Optical Feedback Imaging) to coherent microscopy is presented. This simple and efficient technique using frequency-shifted optical feedback needs the sample to be scanned in order to obtain an image. The effects on magnitude and phase signals such as vignetting and field curvature occasioned by the scanning with galvanometric mirrors are discussed. A simple monitoring method based on phase images is proposed to find the optimal position of the scanner. Finally, some experimental results illustrating this technique are presented.

## *1. Introduction*

The coherent re-injection of light in a laser cavity can be at the origin of a modification of the laser behavior. Often considered as a parasitic phenomenon, it can for example drive the laser in a chaotic regime [1]. When it is correctly controlled, the optical re-injection in a laser can however be used in order to improve its characteristics, such as for example the linewidth narrowing [2], or for metrology purpose [3-10], thus conferring on the laser the role of source and detector at the same time. Since 1963, this principle was employed in an autodyne configuration to measure distances or speeds [11,12]. More recently, a heterodyne imagery technique called LOFI (Laser Optical Feedback Imaging) was developed [13] and has been used for miscellaneous applications [14-16]. Coherent microscopy is one of many areas of application of this versatile technique. It provides access to the complex amplitude of the backscattered field and thus allows obtaining a high resolution, 3D representation of a sample [17-19]. Classically, to measure the height of a sample with a high resolution (i.e. much smaller than the wavelength), one have to build an interferometer which is usually complicated to implement. The LOFI technique allows making interferometric measurement with a particularly simple arrangement which can easily be adapted on a conventional microscope to provide an existing setup with new imaging modalities. Moreover, under some particular conditions, the reinjection of light in the laser may introduce an optical amplification of the interferometric signal.

---

[1] Present address : Ecole Normale Supérieure, Laboratoire Kastler Brossel, Paris, France.

## *2. Principle*

The LOFI technique allows the coherent detection of the photons that have been backscattered by a target [20]. For that purpose, a laser which is a coherent light source is used in a heterodyne interferometer of which the reference arm is replaced by the laser cavity itself. By doing so, interferences occur inside the laser cavity because of the reinjection of a small amount of the backscattered light, accordingly to the reverse path principle. These interferences can be analyzed by measuring a fraction of the laser output power. This approach has several advantages that will be described below.

First, the detection of backscattered photons allows working with a very wide range of targets, transparent or opaque, thin or thick. Only a single interface between two media with different optical index is needed to get a signal.

Then, much information is contained in the complex amplitude of the interferometric signal. Its magnitude depends on the amount of detected photons, which is linked to the refractive index contrast, absorption and diffusion coefficients, polarization and even the target topology. On the other hand, its argument is governed by the optical path length between the laser and the target. This quantity depends on the refractive index and thicknesses of the media through which light traveled. Several kinds of images can then be realized, with a contrast generated by the opto-geometrical parameters of the target, thus having different significations and values. The coherent detection is also favorable to obtain high resolution images in a scattering medium [26], because the multi-scattered photons have lost their coherence. Only the ballistic photons that keep the spatial information of their interaction with the target are detected. Unfortunately, they are usually very few and it is therefore necessary to implement a very sensitive detection scheme.

The use of the laser cavity as a reference also brings some benefits. First, there is no need for a delicate alignment between the probe beam and the reference beam because the setup is self-aligned by construction. Then, the main trump of this technique lies in the interferometric signal amplification by a resonance phenomenon in the laser dynamics. Such amplification is made possible by the use of a class B laser which is characterized by a photon lifetime in the cavity that is shorter than the fluorescence lifetime of the active medium. These lasers show relaxation oscillations of their intensity and are very sensitive to optical feedback, particularly when the reinjected light has a frequency shift which is close to the relaxation frequency. We have shown that this gain increases the signal to noise ratio of the system for given laser power and noise detection level [21,22].

## *3. Experimental setup*

Figure 1 shows a schematic of the experiment. In this study, we used a diode-pumped $Nd^{3+}$:YAG microchip laser with a 800 µm thick cavity which emits a radiation of a few mW at the wavelength of 1.064 µm. Its relaxation frequency is around 1 MHz. The heterodynation is realized with a frequency shifter constituted by two acousto-optic deflectors (IntraAction, ATM-801A2), respectively operating at 81.5 MHz (order +1) and 81.5 MHz-$\Delta f$ (order –1), so that the overall frequency shift is $\Delta f$. This frequency shift can be controlled from the AOD driver (IntraAction, DFE-804A4A). It is chosen close to the relaxation frequency of the laser in order to profit from the dynamical amplification phenomenon. A two-axis galvanometric mirror scanner (Cambridge Technology, 6220) makes it possible to move the beam on the surface of the sample to obtain an image. The beam is injected into the microscope (Zeiss, Axio Imager M1) via the side port and is focused on the sample by the objective (Zeiss, A-Plan 40x/0.65). Because of the great sensitivity of the system with respect to any back-reflection, all the lenses (except for the microscope objective) are anti-reflection coated for the working wavelength: 1.064 µm. A fraction of the output intensity of the laser is sent on a Si-photodiode with a beam splitter. The magnitude and phase of the interferometric signal are provided by a lock-in amplifier operating at 2Δf (Stanford Research, SR844) and digitized by a 12 bits data acquisition board (National Instruments, NI USB-6211). Home-made software developed under the LabVIEW environment has in charge to drive the scanner and to synchronously acquire the data in order to build the images.

## *4. Scanner induced distortions*

The phase measurements are those that require the most attention. In order to have good quality images, it is necessary to scan the target quickly to avoid any drift during the acquisition. One should also remove any source of vibrations. For these reasons, it is preferable to use a galvanometric mirrors scanner to scan the beam on the target rather than moving the target itself. However, this solution also has some drawbacks that result in distorted magnitude and phase images.

In an ideal case, the scanner must be positioned in a plane conjugate to the entrance pupil of the microscope objective to avoid the vignetting phenomenon when scanning. In practice, it is not possible to satisfy this condition simultaneously for both the scanner mirrors (unless they are conjugated with additional optics, which complicates the setup and makes it more cumbersome). Therefore, it occasions a vignetting effect that can be easily assessed.

The last part of the setup taking place between the scanner and the target is detailed on Figure 2. $P_1$ is the conjugate plane of the entrance pupil Pu of the microscope objective $L_5$. The scanner mirrors $M_1$ and $M_2$ are separated by a distance e. They are arranged on either side of the plane $P_1$ so that $M_1$ is separated from P by a distance d. For simplicity of representation, the deflection by the two mirrors has been represented in the same plane. From this figure, it is easy to establish that when the $M_1$ mirror deflects the beam by a small angle α from the optical axis, the lateral shift of the beam compared to the entrance pupil of the objective is:

$$|P'S| = \frac{f_4}{f_3} d\alpha \qquad (1)$$

In the same way, we have for the mirror $M_2$:

$$|P'T| = \frac{f_4}{f_3} (e-d)\alpha \qquad (2)$$

$L_3$ and $L_4$ lenses are wide enough to avoid losses during the scan. In order to profit from the whole numerical aperture of the microscope objective $L_5$ and, in the same time, to limit the power losses, the lenses system has been designed so that the radius of the beam, that is assumed to be gaussian (see Figure 3), is 1.5 times smaller than the radius of the objective entrance pupil, corresponding to a 99% power transmission ratio when α=0°. By computing the ratio of the field that passes through the objective pupil as a function of the lateral shift, one can then quantify the vignetting effect on magnitude images of a flat and uniformly scattering object located in the observation plane $P_3$ (Figure 4). These results suggest that the best solution to minimize the vignetting effect on magnitude images (therefore to limit the degradation of the signal to noise ratio) is to share it out on both dimensions by locating the scanner so that the ideal plane $P_1$ is exactly at the same distance from the mirrors. By doing so, the magnitude is significantly affected only on the edge of the accessible field. The resulting magnitude variations can then easily be corrected with a post processing of the images.

On the other hand, the length of the optical path between the laser cavity and the laser spot on the target is modified during the scan. This effect can also be calculated. The path length variations in both dimensions will be determined by comparison with a reference path, in dotted line on Figure 2. Whatever the value of the deflection angle α of the beam relative to the optical axis, all the optical paths between P and P' have the same length because these points are conjugated. Moreover, because Pu and $P_3$ are in the focal planes of $L_5$, P' is at the same optical distance from every point on $P_3$. The reference optical path length is then independent of α.

If the first scanner mirror $M_1$ is at a distance d from P, it is easy to find that the difference between the real and the reference optical paths is given by :

$$|M_1Q| - |M_1P| = d(\cos(\alpha) - 1) \quad (3)$$

Similarly, the distance between the mirror $M_2$ and the point P is responsible for an optical path length variation when scanning which is equal to:

$$|PM_2| - |PR| = (e - d)(1 - \cos(\alpha)) \quad (4)$$

By considering that the mirrors $M_1$ and $M_2$ deflect the beam by an angle α and β respectively, the phase function of the scanning process can be determined from equations (3) and (4):

$$\Phi(\alpha, \beta) = \Phi_0 + 2k[d(\cos(\alpha) - 1) + (e - d)(1 - \cos(\beta))] \quad (5)$$

where $\Phi_0$ is an arbitrary phase term, $k = 2\pi/\lambda$ is the wave number and the factor 2 outside the brackets corresponds to a round trip between the laser and the target. Equation (5) has been evaluated for three different scanner positions and the results are shown on Figure 5. These images correspond to the phase images that would be obtained by scanning a flat target which is perpendicular to the optical axis. To recover an image that corresponds to the target profile, the phase must then be processed to remove the scanner induced field curvature.

Finally, another effect will be discussed. So far, we have assumed that the mirrors were hit by the beam on their rotation axis, which is not obvious to achieve in practice without monitoring. According to the Figure 6, when the beam is correctly centered on the mirror (i.e. the optical axis intersects the rotation axis) it is reflected at point A, along the dotted line, for any value of α. If it is off-center, then the mirror is hit at point B after a rotation through an angle α/2, introducing an optical path length difference, in comparison with the centered case, which is given by:

$$|AB| + |BC| = g\left[\frac{1}{\sqrt{2}} - \sin\left(\frac{3\pi}{4} + \alpha\right)\right] \quad (6)$$

During the scan, an additional phase variation is thus introduced by this decentering:

$$\Phi_g(\alpha) = 2kg\left[\frac{1}{\sqrt{2}} - \sin\left(\frac{3\pi}{4} + \alpha\right)\right] \quad (7)$$

As can be seen on Figure 7, the correction is noticeable, even for a small shift. A similar expression of this correction is obtained for each mirror, which allows us to compute a new phase pattern taking into account the effect of mirrors decentering (Figure 8). It should be noted that a bad centering is also responsible for an additional vignetting effect on magnitude images.

To summarize, the monitoring of the phase images is a good and convenient way to adjust the position and the centering of the scanner. The best configuration which minimizes the vignetting effect corresponds to a centered, saddle shaped phase image.

## *5. Phase unwrapping*

When scanning a target, the phase signal provided by the lock-in amplifier is wrapped, which means that it is known modulo $2\pi$. To recover the real profile of a target, without discontinuity, the phase signal must still be unwrapped. The MLBT algorithm [23] has been chosen for its robustness towards experimental noise. However, correct phase unwrapping is impossible if the phase fringes are not sufficiently spatially resolved and aberrant results can then be obtained. It is therefore not advisable to unwrap the raw phase image because, as we have shown in the previous section, the scanner induced phase variations are very fast away from the optical axis (i.e. away from the image centre) and it would necessitate a large spatial oversampling which could be completely needless, considering the spatial resolution of the image. Therefore, it is wiser to subtract modulo $2\pi$ the scanner phase function from the raw phase image before unwrapping it.

## *6. Experimental results*

The LOFI microscopy technique was implemented to get magnitude and phase images on two different kinds of samples. The first one is a part of an AFM calibration grid constituted by a set of microscopic blocks etched on a silicon surface. This sample was used first in order to calibrate the horizontal scales of our images. On the magnitude image (Figure 9), the surface micro-structure is clearly visible. The lateral resolution is given by the beam waist on the target and is about 0.5 µm with the microscope objective that has been used. This value is confirmed by the thickness of the blocks edges (dark lines on the image). The image has been voluntarily oversampled in order to facilitate the blocks edges localization with a precision that is better than the optical resolution. Since the whole sample is made of silicon, the image contrast is obviously not only governed by the optical index contrast between the sample and the surrounding medium (the air in this case), but also by the geometrical characteristics of the sample surface. For instance, there is a contrast between the blocks and the space around them because these regions have a different altitude, and we know that the detection efficiency highly depends on the distance between the scattering/reflecting surface and the focal plane [24]. In the case of a reflecting surface, the incidence angle of the beam also determines the amount of light that will be reinjected, hence the signal magnitude (the higher magnitude is obtained when the beam has a normal incidence). This is the reason why the blocks borders appear darker, as the image edges where the incident beam is at an angle because of the scanning process (see Figure 2). It should be noted that the vignetting effect is negligible in this case because of the small dimensions of the image compared to the accessible field size.

On the other hand, the phase image (Figure 10) shows the true topography of the sample with an interferometric resolution [25], that is to say about a few nm on the vertical scale. According to the manufacturer, the blocks height is 140 nm ± 10%. The precision of the phase determination is strongly dependant on the signal to noise ratio, which is enhanced by the LOFI technique. Thus, it is possible to have a good quality phase measurement, even with a weak backscattered signal (which is typically the case on the blocks borders). Magnitude and phase images have been acquired simultaneously in 10 s for a size of $256 \times 256$ pixels. This time corresponds to 150 µs by pixel which is the response time of our instrumentation.

The other object that we have observed is a biological sample. The cells from a fresh blood drop have been separated on a glass microscope slide by the classical blood smear technique. The edge of a red blood cell is visible on the magnitude image (Figure 11) but the phase image (Figure 12) is much more interesting as we clearly see the characteristic concave shape of the cell. This image also shows that the cell seems to be surrounded by some substance, probably the plasma. It should be noted that there is no height scale on Figure 12. Indeed, when dealing with thin and transparent objects such as a cell in the near infrared, the measured phase is no longer proportional to the object height. A very simple model can offer an explanation to this phenomenon. It results from the interference between the signals scattered or reflected by the different interfaces of the object when they are very close (that means separated by a distance which is small or comparable to the Rayleigh range). Let us consider a target constituted by a thin transparent layer on a semi-infinite substrate (Figure 13).

The first contribution to the detected signal comes from the reflection at the superstrate/layer interface. In the reference plane, after a round trip, the complex amplitude of the reflected field is proportional to:

$$\mathbf{E_1} = E_1 \exp(i\Phi_1) \tag{8}$$

with

$$E_1 = E_0 \left( \frac{n - n_{sup}}{n + n_{sup}} \right) \tag{9}$$

and

$$\Phi_1 = 2k(d-h)n_{sup} \tag{10}$$

where $E_0$ is the incident field magnitude and $k=2\pi/\lambda$ is the wave number.

The second contribution comes from the layer/substrate reflection (one reflection on this interface and two transmissions through the other interface):

$$\mathbf{E_2} = E_2 \exp(i\Phi_2) \tag{11}$$

with

$$E_2 = E_0 \left( \frac{2n_{sup}}{n + n_{sup}} \right) \left( \frac{n_{sub} - n}{n_{sub} + n} \right) \left( \frac{2n}{n + n_{sup}} \right) \tag{12}$$

and

$$\Phi_2 = 2k[(d-h)n_{sup} + hn] \tag{13}$$

One then detects the coherent sum of these two fields, whose magnitude and phase have been plotted on Figure 14 as a function of the layer parameters n and h, in the particular case where $n_{sub}=1.51$ and $n_{sup}=1$ corresponding to a sample on a microscope slide in air. It is clear that the phase is not proportional to the layer thickness, except for $n=n_{sub}$ (like in Figure 10) because in that particular case, the layer/substrate interface vanishes and there is no more interference.

However, Figure 14 suggests that it is possible to simultaneously determine the thickness and the optical index of the layer from both magnitude and phase of the reflected field. Unfortunately, it only works in normal incidence; therefore, this method can't be applied for profilometry purpose because the detected magnitude also depends on the interfaces locale orientations [24] which are generally unknown.

## *7. Conclusion and outlook*

The application of the LOFI technique to coherent microscopy was presented. The simple setup can easily be implemented on an existing microscope in order to extend its imaging modalities.

The LOFI is a point based technique; it means that to build an image, the target must be scanned. A galvanometric mirrors scanner moving the beam on the sample surface is a fast and convenient solution, but it induces some distortions on images that have been discussed. In particular, the phase pattern generated by the scanning process can be useful to optimize the scanner positioning in order to reduce the vignetting effect.

Finally, magnitude and phase images obtained on a micro-structured silicon sample and a red blood cell have been presented to illustrate the potentialities of this non conventional imaging technique.

These results encourage us to develop this technique in order to realize in depth microscopy of scattering media such as biological tissues. Indeed, the characteristics of the LOFI are particularly interesting for the biomedical applications. The used laser emits a radiation of a few mW at 1.064 µm that suits well for biological tissues observation. Coherent detection with the great sensitivity provided by the dynamical amplification performed by the used laser makes it possible to obtain images in scattering media, without need for contrast agent [26]. In addition, the TEM$_{00}$ mode of the laser plays the role of a spatial filter during the detection stage, like the pinhole of a confocal microscope, which allows the localization of the signal in the three dimensions of space [27]. However, some work still has to be done to eliminate the spurious reflections that scramble the image by causing interferences when they are superimposed on the useful signal from the focal plane. Usually negligible, the spurious reflections start to be annoying when the signal is very weak, which is typically the case when it travels trough a scattering medium. The most important contributions to this parasitic signal come from the microscope objective and the surface of the sample. The presence of a cover plate on the sample also causes an important unwanted echo. A promising solution based on a multi-beam illumination of the sample [28] is currently being implemented.

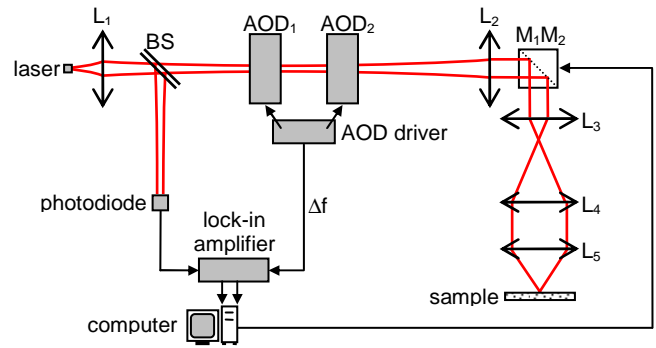

**Figure 1** – Description of the experimental set-up: $L_1$, collimation lens; BS, beam splitter; $AOD_1$, acousto-optic deflector at 81.5 MHz (order +1); $AOD_2$, acousto-optic deflector at 81.5 MHz-$\Delta f$ (order –1); $L_2$, collimation lens; $M_1M_2$, two-axis galvanometric mirrors scanner; $L_3$, scan lens; $L_4$, tube lens; $L_5$, microscope objective.

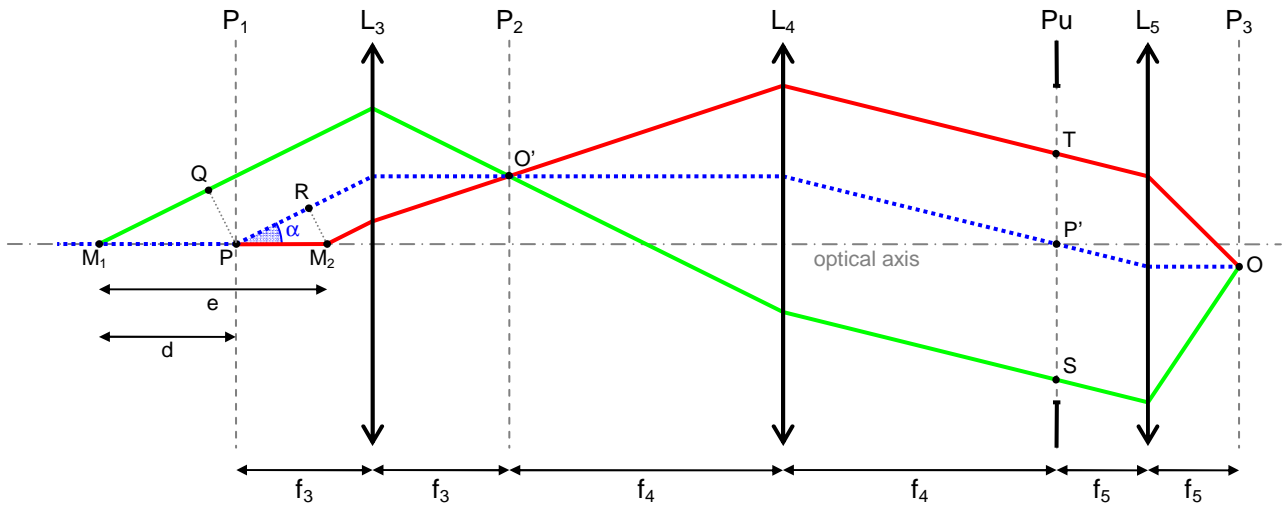

**Figure 2** – Ray tracing for the computation of amplitude and phase images distortions. $f_3$, $f_4$ and $f_5$ are respectively the focal lengths of the $L_3$, $L_4$ and $L_5$ lenses. Pu is the entrance pupil of $L_5$ and $P_3$ is the observation plane.

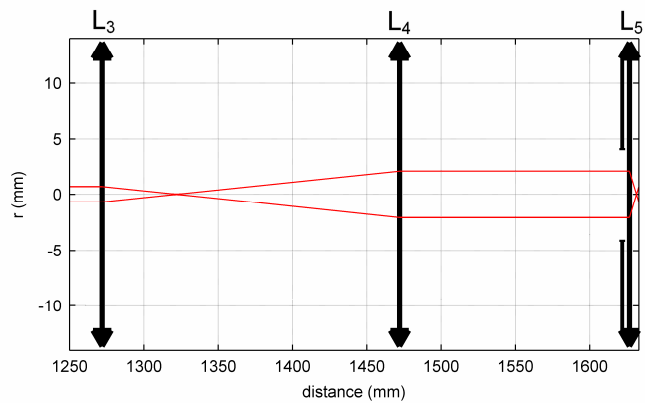

**Figure 3** – 1/e amplitude contours of the gaussian beam between the scanner and the observation plane. The parameters for the simulations are: $\alpha=0°$; $f_3=50$ mm; $f_4=150$ mm; $f_5=4.7$ mm.

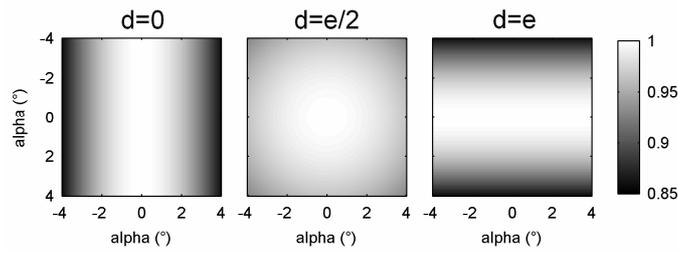

**Figure 4 – Computed magnitude images exhibiting the vignetting effect for three different positions of the scanner. The parameters for the simulations are: e=9.1 mm; $f_3$=50 mm; $f_4$=150 mm; objective pupil radius=4 mm; beam radius on the objective pupil=2.66 mm.**

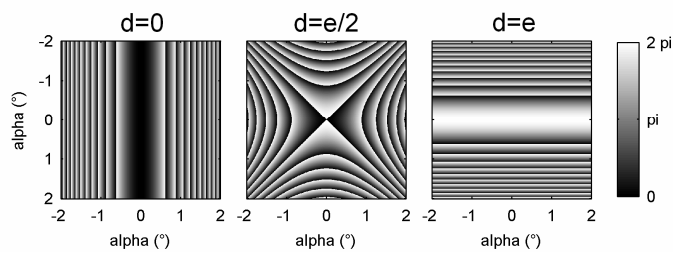

**Figure 5 – Computed phase patterns generated by the scanning process for three different positions of the scanner. The parameters for the simulations are: $\varphi_0$=0 rad; $\lambda$=1064 nm; e=9.1 mm.**

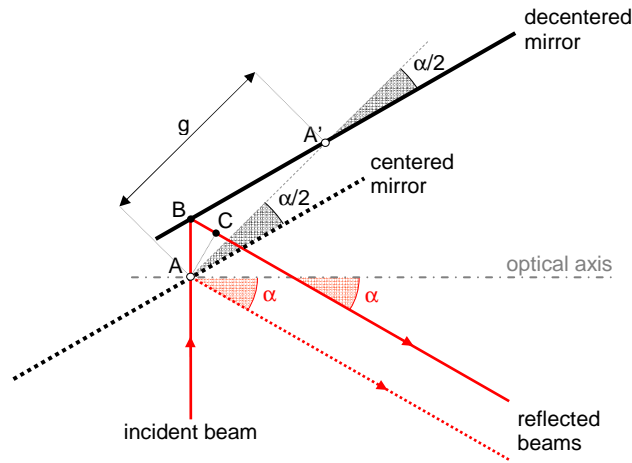

**Figure 6 – Rays reflected by one of the scanner mirrors when the beam intersects (dotted line) or not (plain line) its rotation axis. A and A' are respectively the rotation axis of the centered and decentered mirrors, separated by a distance g; $\alpha$=0° corresponds to the incident beam reflected along the optical axis in both cases.**

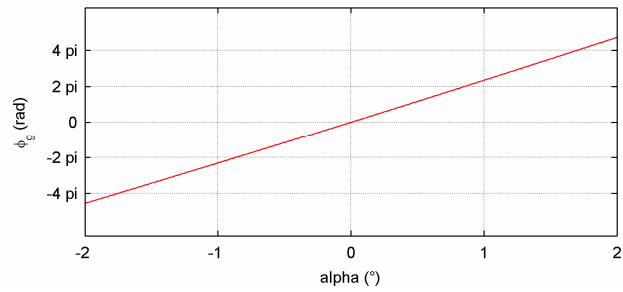

**Figure 7** – Additional phase distortion along the direction scanned by the mirror whose rotation axis is shifted. The parameters for the simulations are: $\lambda$=1064 nm; e=9.1 mm; d=4.45 mm; g=50 µm.

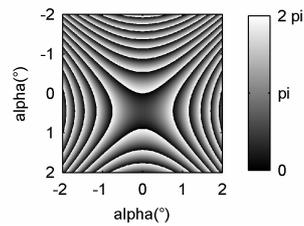

**Figure 8** – Shifted phase pattern computed when the laser beam doesn't intersect one of the mirrors rotation axis. The parameters for the simulations are: $\lambda$=1064 nm; e=9.1 mm; d=e/2; $g_{M1}$=50 µm; $g_{M2}$=0 µm; -2°≤$\alpha$,$\beta$≤2°.

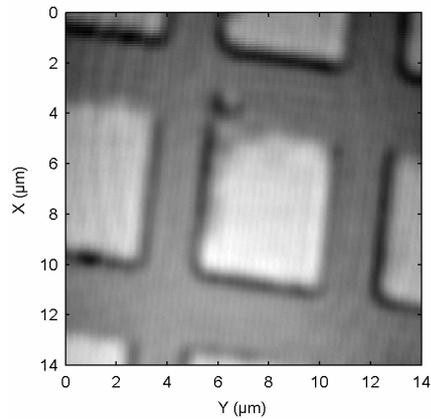

**Figure 9** – Magnitude image of a micro-structured silicon surface (256×256 pixels).

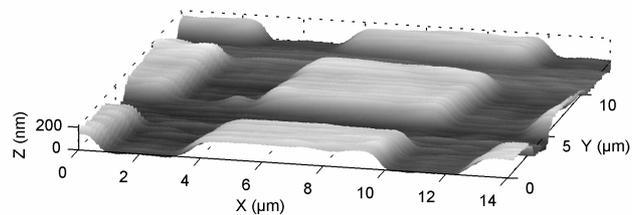

**Figure 10** – Phase image of a micro-structured silicon surface (256×256 pixels).

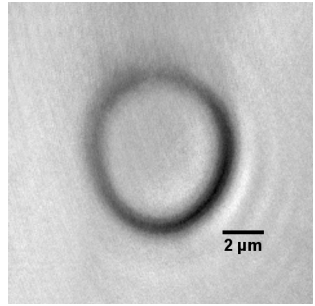

**Figure 11 – Magnitude image of an isolated red blood cell on a glass slide (200×200 pixels).**

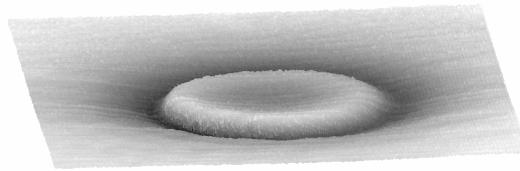

**Figure 12 – Phase image of an isolated red blood cell on a glass slide (256×256 pixels).**

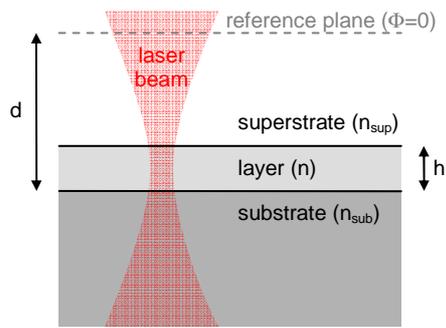

**Figure 13 – Model for the multiple scattering/reflection effect assessment.**

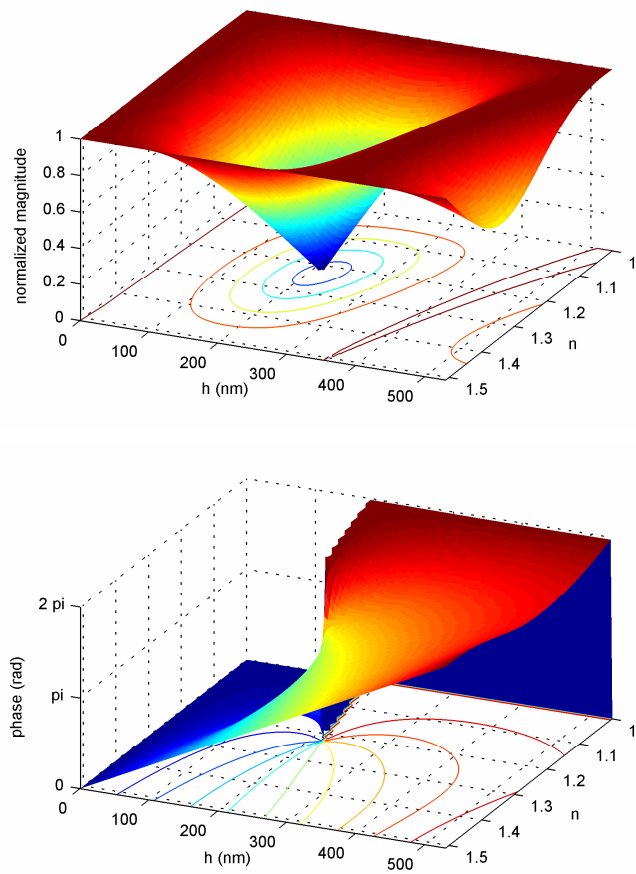

**Figure 14 – Normalized magnitude and phase of the field reflected by the structure on Figure 13 when $n_{sub}$=1.51 and $n_{sup}$=1.**